\documentclass[         %
aps,                    %  American Physical Society
prd,                    %  Physical Review L
showpacs,               %  Displays PACS after abstract
superscriptaddress,    %  Authors' addresses linked with superscripts
nofootinbib,            %  Does not treat footnotes as references
twocolumn,             %
showkeys,               %
preprintnumbers,        %
a4paper,               %  A4 paper format
floatfix]               %  Fixes float errors
{revtex4-1}               %  REVTEX 4 Package used
\usepackage{graphicx,amsmath}
\usepackage{mathrsfs,bm} %%% <<--- math fonts
\usepackage{color}
\usepackage{graphicx}
\usepackage{multirow}
\usepackage{hyperref}
\usepackage{nicefrac}
\usepackage[utf8]{inputenc}
\usepackage[caption=false]{subfig}
\usepackage{feynmp}
\usepackage{soul}
\newcommand{\f}{\uparrow\downarrow\mkern-23mu\textrm{---}}
\newcommand{\e}{\textrm{---}}
\newcommand{\tf}{\parbox{4mm}{$\uparrow\downarrow\mkern-23mu\textrm{---}$}}

\makeatletter
\g@addto@macro\normalsize{%
  \setlength\abovedisplayskip{7pt}
  \setlength\belowdisplayskip{7pt}
  \setlength\abovedisplayshortskip{5pt}
  \setlength\belowdisplayshortskip{5pt}
}
\makeatother

\newcommand{\msp}{\hspace{0.2pt}}

\newcommand{\s}{$\mkern5mu$} %%% <<---  use ~\cite to avoid line breaks

\begin{document}
%%-----------------------------------------------------------------%
\title{Spectral splits of neutrinos as a BCS-BEC crossover type phenomenon}
%%-----------------------------------------------------------------%
\author{Y. Pehlivan}
\email{yamac.pehlivan@msgsu.edu.tr}
\affiliation{Department of Physics, Mimar Sinan Fine Arts University, Sisli, Istanbul, 34380, Turkey}
%%-----------------------------------------------------------------%
\author{A. L. Suba{\c s}{\i}}
%\email{alsubasi@itu.edu.tr}
\affiliation{Department of Physics, Istanbul Technical University, Istanbul, Turkey}
%%-----------------------------------------------------------------%
\author{N. Ghazanfari}
%\email{nadirgh@gmail.com}
\affiliation{Department of Physics, Mimar Sinan Fine Arts University, Sisli, Istanbul, 34380, Turkey}
%%-----------------------------------------------------------------%
\author{S. Birol}
%\email{savas.birol@gmail.com}
\affiliation{Department of Physics, Istanbul University, Istanbul, Turkey}
%%-----------------------------------------------------------------%
\author{H. Y\"{u}ksel}
%\email{hyuksel@gmail.com}
\affiliation{Department of Physics, Mimar Sinan Fine Arts University, Sisli, Istanbul, 34380, Turkey}
%%-----------------------------------------------------------------%
\date{\today}
%%-----------------------------------------------------------------%
\begin{abstract}
We show that the spectral split of a neutrino ensemble which initially consists
of electron type neutrinos, is analogous to the BCS-BEC crossover already
observed in ultra cold atomic gas experiments. Such a neutrino ensemble mimics
the deleptonization burst of a core collapse supernova. Although these two
phenomena belong to very different domains of physics, the propagation of
neutrinos from highly interacting inner regions of the supernova to the vacuum
is reminiscent of the evolution of Cooper pairs between weak and strong
interaction regimes during the crossover. The Hamiltonians and the corresponding
many-body states undergo very similar transformations if one replaces the pair
quasispin of the latter with the neutrino isospin of the former.
\end{abstract}
%%-----------------------------------------------------------------%
\medskip
\preprint{}
\maketitle
%%------------------------------------------------------------------%

%%%%%%%%%%%%%%%%%%%%%%%%%%%%%%%%%%%%%%%%%%%%%%%%%%%%%%%%%%%%%%%%%%%%%%%%%%%
\section{Introduction}

A core-collapse supernova releases $99\%$ of its energy in the form of neutrinos
in the MeV energy scale~\cite{Colgate:1966ax,Woosley:1986ta}.  Our basic
understanding about these neutrinos was
confirmed~\cite{1989ARA&A..27..629A,Burrows:1987zz} when supernova 1987A
exploded in our neighbor galaxy, the Large Magellanic Cloud, and generated $19$
neutrino events in Kamiokande~\cite{Hirata:1987hu} and IBM~\cite{Bionta:1987qt}
detectors. The next important breakthrough in this field will be the observation
of neutrinos from a supernova explosion in our own galaxy which is estimated to
generate thousands of neutrino events in current neutrino
detectors~\cite{Scholberg:2012id}.  Therefore a future galactic supernova
presents a unique opportunity to test our understanding of neutrinos.  This
includes the many-body aspects of their flavor transformations~\cite{Pantaleone:1992xh,
Pantaleone:1992eq} which develop via the neutrino-neutrino ($\nu\nu$)
interactions in the supernova~\cite{Bardin:1970wq,fuller&mayle}.

\begin{figure}[t]
%\vspace*{-0.cm}
%\includegraphics[width=0.95 \columnwidth,clip=true]{fig1}
%
%\vspace*{-0.cm}
\subfloat[\label{nu-nu forward}]{
\includegraphics[width=0.46 \columnwidth,clip=true]{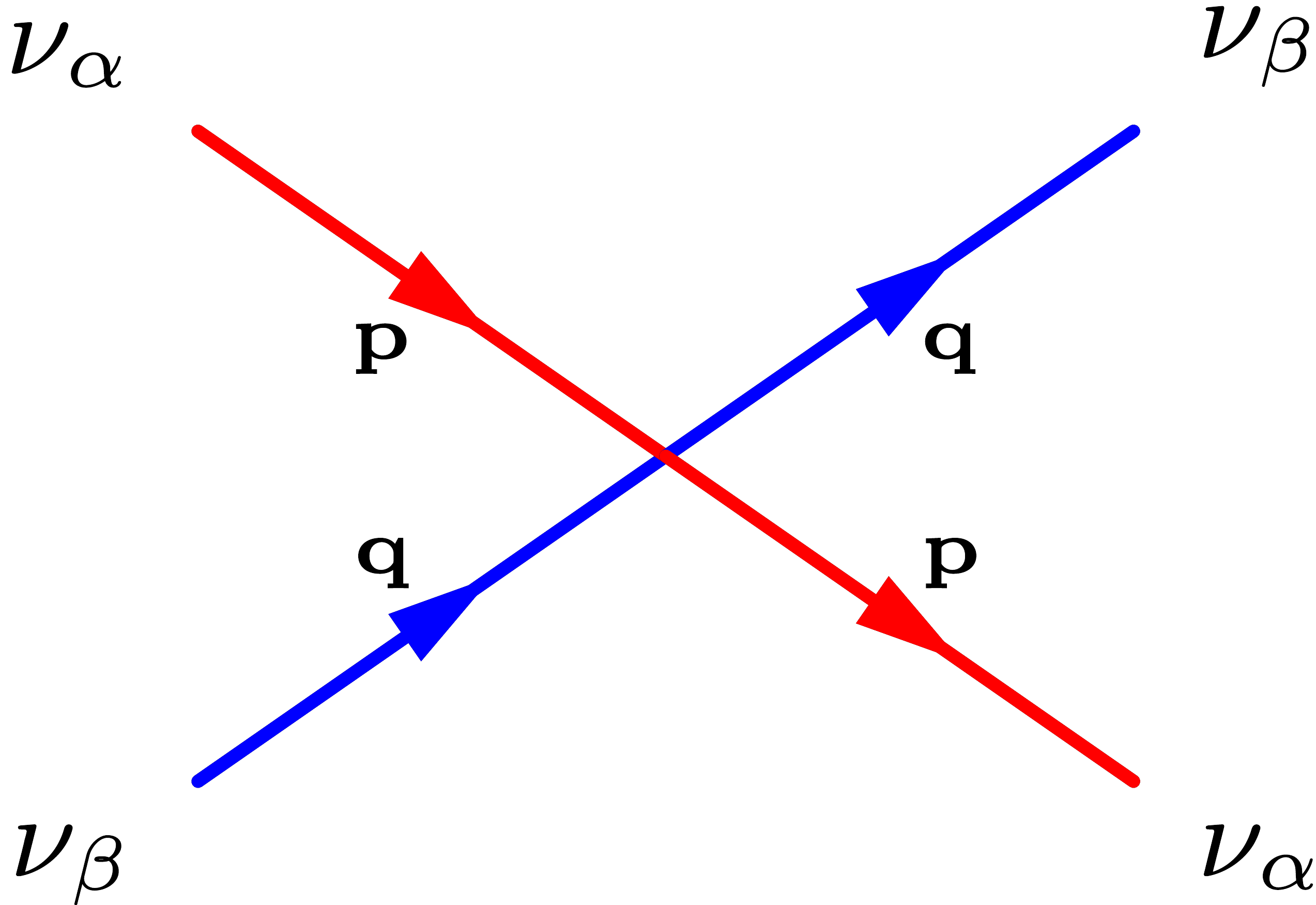} \vspace*{-0.3cm} 
}
\hspace*{0.3cm}
\subfloat[\label{nu-nu exchange}]{
\includegraphics[width=0.46 \columnwidth,clip=true]{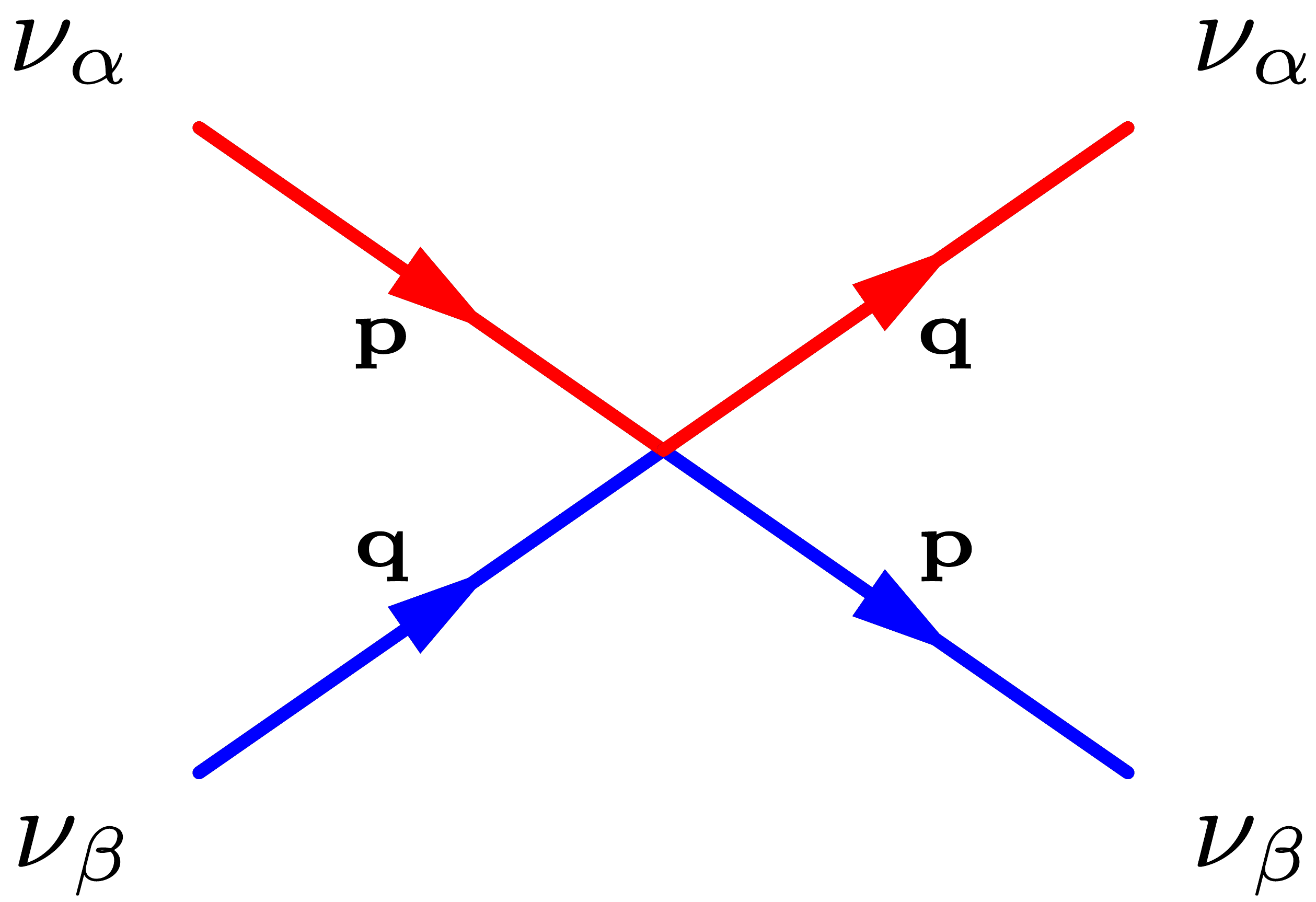} \vspace*{-0.3cm}
}
%
%\vspace*{-3mm}
\caption{(Color online) Forward (a) and exchange (b) diagrams which add up coherently in $\nu\nu$ scattering.
\label{nu-nu}}
%\vspace*{-9mm}
\end{figure}

Although neutrino cross sections are extremely small, their tiny scattering
amplitudes can add up coherently to give rise to a finite effect when neutrinos
propagate in the presence of a matter background~\cite{Wolfenstein:1977ue}.
This is similar to the refraction of light in matter except that, since
neutrinos can interact with each other via neutral current, they can also create
a \emph{self refraction} effect on themselves~\cite{fuller&mayle}. Two kinds of
diagrams, shown in Fig.\s\ref{nu-nu}, add up coherently in self refraction:
(a) the forward scattering diagram in which there is no momentum transfer between particles
and (b) the exchange diagram in which particles completely swap their
momenta~\cite{Pantaleone:1992xh, Pantaleone:1992eq}. The former gives rise to an ordinary refraction index
through the optical theorem~\cite{Wolfenstein:1977ue}.  The latter can be viewed
as a \emph{flavor-exchange} diagram between neutrinos and, as such, it couples
the flavor transformation of each neutrino to the flavor content of the entire
neutrino ensemble.  This turns the flavor evolution of neutrinos near the core
of a supernova into a many-body problem~\cite{Sawyer:2005jk, Balantekin:2006tg,
Pehlivan:2011hp, Pehlivan:2014zua, Volpe:2013jgr, Serreau:2014cfa}.

%%%%%%%%%%%%%%%%%%%%%%%%%%%%%%%%%%%
%%%%%%%%%%%%%%%%%%%%%%%%%%%%%%%%%%%

The correlations between flavor histories of neutrinos with different energies,
which are referred to as \emph{collective neutrino oscillations}, have been
extensively studied~\cite{Samuel:1993uw, Kostelecky:1994dt, Samuel:1995ri,
Duan:2006an, Friedland:2010sc,Duan:2010bg,Mirizzi:2015eza,Chakraborty:2016yeg}.
The large array of resulting nonlinear and emergent behavior displayed by self interacting neutrinos
are reminiscent of condensed matter systems.  A formal analogy between
collective neutrino oscillations and BCS pairing model of
superconductivity~\cite{Bardeen:1957mv} has recently been pointed out by
Pehlivan \emph{et al.}~\cite{Pehlivan:2011hp,Pehlivan:2014zua} and further
elaborated in~\cite{Baker:2016ohg}. Besides the Cooper pairs of electrons in
superconductors, BCS pairing is observed in a broad range of many body systems,
including neutron stars and atomic nuclei~\cite{RevModPhys.75.607}, ultra cold
atomic gases~\cite{RevModPhys.80.885,RevModPhys.80.1215} and excitonic
condensates in semiconductor structures~\cite{PhysRevLett.74.1633,
PhysRevB.72.115320, Byrnes2014}.

%%%%%%%%%%%%%%%%%%%%%%%%%%%%%%%%%%%%%%%%%%%%%%%%%%%%%%%%%%%%%%%%%%%%%%%%%%%
\begin{figure}[t]
%\vspace*{5mm}
	%\includegraphics[width=.9\columnwidth]{../illustration/illustration.png}
	\includegraphics[width=.9\columnwidth]{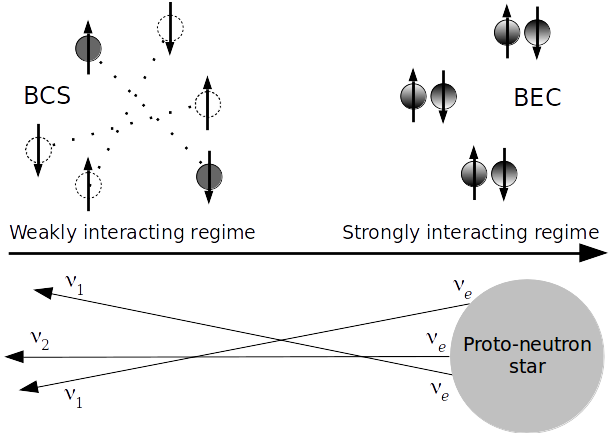}
	%\includegraphics[scale=0.3]{../illustration/illustration}
%\captionsetup{justification=RaggedRight}
	\caption{\small The correspondence between self interacting neutrinos in a core 
		collapse supernova and BCS-BEC crossover. Neutrinos are emitted by
		the proto-neutron star which forms at the center after the core collapse.
		Just outside the surface of the proto-neutron star, neutrino self interaction rate is very 
		high which corresponds to the BEC limit. As the neutrinos move away 
		from the center, the interaction rate decreases and approaches the 
		BCS limit.
	\label{fig:illustration}}
%\vspace*{-5mm}
\end{figure}
%%%%%%%%%%%%%%%%%%%%%%%%%%%%%%%%%%%%%%%%%%%%%%%%%%%%%%%%%%%%%%%%%%%%%%%%%%%
One collective behavior observed in some numerical simulations of neutrinos
emerging from supernova is the \emph{spectral split} or \emph{spectral swap}
phenomenon in which neutrinos in different flavor (or mass) eigenstates
completely exchange their spectra around a certain critical
energy~\cite{Duan:2006an,Duan:2010bg}.  In this paper we show that, for a neutrino
ensemble which initially consists of only electron type neutrinos,
the formation of the spectral split also corresponds to
the well known BCS-BEC crossover~\cite{Zwierlein, Randeria2014} phenomenon.
%as illustrated in Fig.~\ref{fig:illustration}. 
We describe the neutrinos using the effective
two flavor mixing scenario and the neutrino bulb model under the mean field
approximation. Such a model can be considered as an heuristic description of the
initial deleptonization phase of a core collapse supernova.
We illustrate BCS-BEC crossover correspondence in ultra cold
atomic gases which have been used to simulate other quantum 
systems as in creation of Dirac monopoles and observation of quantum phase transitions~\cite{Ray:2014sga,Greiner2002}
and make analogies with different physical phenomena like black hole evaporation and cosmological
effects~\cite{PhysRevLett.85.4643, PhysRevLett.94.061302, PhysRevLett.109.220401,Hung13092013}.
The correspondence considered here doesn't imply any pairing correlations
between the neutrinos. As illustrated in Fig.~\ref{fig:illustration}, 
it is a more subtle analogy in
which the electron neutrinos originally released by the proto-neutron star at
the center are identified with the quasi-hole pairs of the BEC (spheres with
graded color) whereas the first and second neutrino mass eigenstates far from
the center are identifed with the real hole (empty spheres) and particle (solid
black spheres) pairs, respectively.

This paper is organized as follows: in Section II, we briefly review the analogy
between BCS pairing Hamiltonian and the self interacting neutrinos. In Section
III, we introduce the analogy between BCS-BEC crossover in cold atoms and the
neutrino spectral split. We discuss our results and conclude in Section IV\@.

%%%%%%%%%%%%%%%%%%%%%%%%%%%%%%%%%%%%%%%%%%%%%%%%%%%%%%%%%%%%%%%%%%%%%%%%%%%

\section{BCS pairs and self interacting neutrinos}
The BCS model describes the superconducting state of weakly interacting fermions
through a coherent superposition of Cooper pairs. In ultra cold dilute gas
systems, the pairing interaction between fermionic atoms can be described by the
mean-field Hamiltonian
\begin{equation}
\label{cold atoms}
H_{\mbox{\tiny BCS}} = \sum_k
\begin{pmatrix}
c_{k\uparrow}^\dagger & c_{\bar{k}\downarrow}^{\phantom{\dagger}}
\end{pmatrix}
\begin{pmatrix}
\epsilon_k - \mu & - g \Delta^- \\
- g \Delta^+ & - \epsilon_k + \mu
\end{pmatrix}
\begin{pmatrix}
c_{k\uparrow}^{\phantom{\dagger}} \\ c_{\bar{k}\downarrow}^\dagger
\end{pmatrix}\msp.
\end{equation}
We assume that the atoms can occupy a discrete set of energy levels $\epsilon_k$
and the operators $c_{k\uparrow}$ and $c_{\bar{k}\downarrow}$ annihilate spin-up
and spin-down fermions, respectively, in the $k^{\mbox{\tiny th}}$ time-reversed
energy levels. The chemical potential $\mu$ is introduced as a Lagrange multiplier to fix
the average particle number. The interactions can be tuned via Fesh-bach resonances~\cite{Inouye1998}
which function as a control knob for the coupling constant $g$. The physics is captured by the mean field
\begin{equation}
\vec{\Delta}=\left(\tfrac{\Delta^++\Delta^-}{2}, \tfrac{\Delta^+-\Delta^-}{2i},
\Delta^0\right)
\end{equation}
where
\begin{equation}
\label{mean field pairs}
\Delta^- =\sum_\ell
\langle c_{\bar{\ell}\downarrow}^{\phantom{\dagger}} c_{\ell\uparrow}^{\phantom{\dagger}} \rangle, \quad
\Delta^0=\sum_\ell
\left \langle\frac{c_{\ell\uparrow}^\dagger
c_{\ell\uparrow}^{\phantom{\dagger}}-c_{\bar{\ell}\downarrow}^{\phantom\dagger}
c_{\bar{\ell}\downarrow}^\dagger}{2} \right \rangle \msp.
\end{equation}
The pairing potential $\Delta^-={\left(\Delta^+\right)}^*$ describes the scattering of zero
center-of-mass momentum pairs and $\Delta^0$ corresponds to the Hartree
potential which
%contributes to the diagonal term. Therefore the Hartree potential
can be included in the definition of $\mu$, i.e.,
$\mu\to\mu+g\Delta^0$.  In Eq.\s(\ref{mean field pairs}), the expectation values
are calculated with respect to a state which satisfies the usual mean field
self-consistency requirements.

The flavor evolution of self interacting neutrinos near the core of a supernova is
described by a mathematically similar Hamiltonian
\begin{equation}
\label{neutrinos}
H_{\nu\nu}\mkern-2mu=\frac{1}{2}\sum_k
\begin{pmatrix}
a_{k2}^\dagger & a_{k1}^\dagger
\end{pmatrix}
\begin{pmatrix}
- \omega_k + \lambda & G P^- \\
G P^+ & \omega_k -\lambda
\end{pmatrix}
\begin{pmatrix}
a_{k2}^{\phantom{\dagger}} \\ a_{k1}^{\phantom{\dagger}}
\end{pmatrix}\msp.
\end{equation}
We assume that neutrinos are box quantized in volume $V$ and therefore occupy a
discrete set of energy levels $\varepsilon_k$. The operator $a_{ki}$ annihilates
a neutrino in the $k^{\mbox{\tiny th}}$ energy level, in the mass eigenstate
with mass $m_i$ ($i=1,2$). To be specific, we assume that $m_1>m_2$ which
corresponds to inverted mass hierarchy.
The oscillation frequency for a neutrino with energy $\varepsilon_k$ associated with this mass difference is given by
\begin{equation}
\omega_k=(m_1^2-m_2^2)/2\varepsilon_k\, .
\end{equation}
%is the oscillation frequency for a neutrino with energy $\varepsilon_k$ associated with this mass difference.
The Lagrange multiplier $\lambda$ plays an
analogous role to the chemical potential $\mu$ in Eq.\s(\ref{cold
atoms}) as discussed in Ref.~\cite{Pehlivan:2011hp}. $\nu\nu$ interaction strength is given by
\begin{equation}
\label{G}
G=\frac{2\sqrt{2}G_F}{V}D(r)
\end{equation}
where the Fermi constant $G_F$ appears due to our use of the Fermi 4-point interaction as
shown in Fig.\s\ref{nu-nu}. We also use the neutrino bulb model which
approximates the angular dependence of the $\nu\nu$
scattering amplitude~\cite{Sigl:1992fn} with an effective geometrical factor $D(r)\propto 1/r^2$
\cite{Duan:2006an} where $r$ is the distance from the supernova center. Combined
with decreasing neutrino density which increases the normalization volume, G
drops as $1/r^4$.

Here, we adopt an effective two flavor mixing scenario
\begin{equation}
\label{neutrino mixing}
a_{ke}^{\phantom{\dagger}}
 \mkern-3mu=\mkern-3mu \cos{\vartheta} a_{k1}^{\phantom{\dagger}}
\mkern-4mu + \mkern-1mu \sin{\vartheta} a_{k2}^{\phantom{\dagger}}\msp,
\mkern9mu
a_{k\mu}^{\phantom{\dagger}}
\mkern-3mu=\mkern-3mu \sin{\vartheta} a_{k1}
\mkern-5mu - \mkern-1mu \cos{\vartheta} a_{k2}^{\phantom{\dagger}}\msp,
\end{equation}
where the effects of the third flavor and the other background particles are
absorbed in a single mixing angle $\vartheta$ and a single mass squared
difference (see e.g.~\cite{Kuo:1989qe,Balantekin:2006tg}). In
Eq.\s(\ref{neutrinos}), $\nu\nu$ interactions
are described by the mean field $\vec{P}$ defined by
\begin{equation}
\label{mean field neutrinos}
P^-=\sum_\ell
\langle a_{\ell 1}^\dagger a_{\ell 2}^{\phantom{\dagger}}\rangle \quad
P^0=\sum_\ell
\left \langle\frac{a_{\ell 2}^\dagger a_{\ell 2}^{\phantom{\dagger}}-a_{\ell 1}^\dagger a_{\ell
1}^{\phantom{\dagger}}}{2} \right\rangle \msp.
\end{equation}
The components $P^\pm$ create the exchange diagrams shown in Fig.~\ref{nu-nu
exchange} while $P^0$ creates the forward scattering diagram in Fig.~\ref{nu-nu
forward}. $P^0$ contributes to the diagonal of Eq.\s(\ref{neutrinos}) and plays
a similar role to that of an Hartree potential for fermionic pairs.  In the case
of neutrinos, $P^0$ is always non-zero, and we include it in the definition of
$\lambda$ in order to highlight the resemblance with the BCS model.

The similarity between Eqs.\s(\ref{cold atoms}) and (\ref{neutrinos})
suggests the mapping
\begin{equation}
\label{mapping}
a_{k2}\leftrightarrow c_{k\uparrow} \quad\mbox{and}\quad
a_{k1}\leftrightarrow c_{\bar{k}\downarrow}^\dagger,
\end{equation}
which reveals the
common $SU(2)$ group structure of these problems. This group is generated by the
\emph{quasispin operators} for BCS pairs~\cite{Anderson58} given by
\begin{equation}
\label{quasispin}
J_k^{-}=c_{\bar{k}\downarrow}^{\phantom{\dagger}} c_{k\uparrow}^{\phantom{\dagger}}
\quad
J_k^0=
\frac{1}{2}\left(c_{k\uparrow}^\dagger
c_{k\uparrow}^{\phantom{\dagger}}-c_{\bar{k}\downarrow}^{\phantom{\dagger}}
c_{\bar{k}\downarrow}^\dagger\right)\msp,
\end{equation}
and the \emph{mass isospin operators} for neutrinos given by
\begin{equation}
\label{isospin}
J_k^{-}=a_{k1}^{\dagger}a_{k2}^{\phantom{\dagger}}
\quad
J_k^0=\frac{1}{2}\left(a_{k2}^\dagger a_{k2}^{\phantom{\dagger}}
-a_{k1}^\dagger a_{k1}^{\phantom{\dagger}}\right)\msp.
\end{equation}
For convenience, we denote both the pair quasispin and the neutrino isospin with
the same symbol but it is always clear which one is being referred to from the
context. In both cases, components of $\vec{J_k}$ obey the $SU(2)$ algebra,
i.e.,
\begin{equation}
[J_k^+,J_\ell^{-}]=2\delta_{k\ell}J_k^0,\quad \mbox{and} \quad
[J_k^0,J_\ell^{\pm}]=\pm\delta_{k\ell}J_k^{\pm}.
\end{equation}

In terms of these operators, the Hamiltonians describing the BCS pairs and
self interacting neutrinos can be written respectively as
\begin{equation} \label{hamiltonians}
\begin{split}
&H_{\mbox{\tiny BCS}}=
\sum_k 2(\epsilon_k-\mu) J^0_k
%-g\left( \langle J^-\rangle J^+ + \langle J^+\rangle J^-\right) \label{hamiltonians}\\
-g\Big( \langle J^-\rangle J^+ + \langle J^+\rangle J^-\Big)\\
&H_{\nu\nu}=
%-\sum_k (\omega_k-\lambda) J^0_k +\frac{G}{2}\left(\langle J^-\rangle J^+ +\langle J^+\rangle J^-\right) \nonumber
-\sum_k (\omega_k-\lambda) J^0_k +\frac{G}{2}\Big(\langle J^-\rangle J^+ +\langle J^+\rangle J^-\Big)
\end{split}
\end{equation}
where $\vec{J}$ is the total quasi-/iso-spin operator for all energy levels,
i.e., $\vec{J}=\sum_k\vec{J}_k$. Note that the two Hamiltonians differ by an
overall minus sign.

Eq.\s(\ref{quasispin}) tells us that for a single energy level $k$, quasispin up
and down states correspond to that level being occupied ($|\tf\rangle$) or
unoccupied ($|\e\rangle$) by a pair, respectively. (Levels occupied by unpaired
atoms decouple from the pairing dynamics and are ignored here.) In the case of
neutrinos, Eq.\s(\ref{isospin}) tells us that the isospin up and down states for
energy level $k$ correspond to the neutrino occupying that energy level being in
$\nu_2$ and $\nu_1$ mass eigenstate, respectively. Therefore, the analogous
states are
\begin{equation}
\label{analog states}
|\f\rangle \leftrightarrow |\nu_2\rangle
\quad\mbox{and}\quad
|\textrm{---}\rangle \leftrightarrow |\nu_1\rangle\msp.
\end{equation}
Considering all the energy levels in the system, the state in which all
neutrinos are in the $\nu_1$ mass eigenstate corresponds to the \emph{particle
vacuum} of the BCS model, i.e., the state with no pairs in it:
\begin{equation}
\label{vacuum}
|\emptyset\rangle \equiv  |\e \; \e \; \e \;  \dots \rangle \leftrightarrow
|\nu_1 \; \nu_1 \; \nu_1 \;  \dots \rangle\msp.
\end{equation}
Clearly the particle vacuum of the BCS model has no dynamics but neither does
the neutrino state on the right hand side of Eq.\s(\ref{vacuum}): since all the
neutrinos are in mass eigenstate, this state does not undergo vacuum
oscillations and since all the neutrinos are in the same mass eigenstate,
exchange diagrams shown in Fig.\s\ref{nu-nu exchange} cannot change this state
either. Acting on both states in Eq.\s(\ref{vacuum}) with $J_k^+$ repeatedly, we
find
\begin{equation}
\label{others}
\begin{split}
&|\f \; \e \; \e \; \e \; \dots \rangle \leftrightarrow
|\nu_2 \; \nu_1 \; \nu_1 \; \nu_1 \; \dots \rangle\msp,
\\
&|\f \; \f \; \e \; \e \; \dots \rangle \leftrightarrow
|\nu_2 \; \nu_2 \; \nu_1 \; \nu_1 \; \dots \rangle\msp,
\end{split}
\end{equation}
and so on. The pairs on the left now scatter between the energy levels
while the neutrinos on the right undergo exchange interactions.

%%%%%%%%%%%%%%%%%%%%%%%%%%%%%%%%%%%%%%%%%%%%%%%%%%%%%%%%%%%
\begin{table}[t]
\begin{center}
\begin{tabular}{r c c l} \toprule
\multicolumn{2}{c}{Self interacting neutrinos}
&
\multicolumn{2}{c}{Fermions with pairing}
\\
\hline
\\[-2ex]
%%%%%
\multirow{2}{*}{Mass eigenstates}
&
$|\nu_1\rangle$
&
$|\e\rangle$
&
\multirow{2}{*}{Pair states}
\\[1ex]
%%%%%
&
$|\nu_2\rangle$
&
$|\f\rangle$
&
\\[3ex]
%%%%%
\multirow{2}{*}{\mbox{Neutrino operators}}
&
$a_{1k}^{\dagger}$
&
$\displaystyle{c_{\bar{k}\downarrow}^{\phantom{\dagger}}}$
&
\multirow{2}{*}{Particle-hole}
\\[1ex]
%%%%%
\mbox{in mass basis}
&
$a_{2k}^{\dagger}$
&
$\displaystyle{c_{k\uparrow}^{\dagger}}$
&
\mbox{operators}
\\[4ex]
%%%%%
\multirow{2}{*}{\mbox{Neutrino operators}}
&
$a_{ek}^{\dagger}$
&
$\displaystyle{\lim_{g\to\infty}\tilde{c}_{\bar{k}\downarrow}^{\phantom{\dagger}}}$
&
\multirow{2}{*}{Quasi particle-}
\\[1ex]
%%%%%
\mbox{in flavor basis}
&
$a_{\mu k}^{\dagger}$
&
$\displaystyle{\lim_{g\to\infty}\tilde{c}_{k\uparrow}^{\dagger}}$
&
\mbox{hole operators}
\\[4ex]
%%%%%
\phantom{\mbox{Neutrino numbers}}
&
\phantom{\multirow{2}{*}{$\displaystyle{\frac{n_1\mkern-1mu(\mkern-1mu k \mkern-1mu) \mkern-3mu -
\mkern-3mu n_2 \mkern-1mu( \mkern-1mu k \mkern-1mu)}{2}}$}}
&
\phantom{\multirow{2}{*}{$\displaystyle{n_p \mkern-1mu ( \mkern-1mu k \mkern-1mu) \mkern-2mu - \mkern-2mu
\frac{1}{2}}$}}
&
\phantom{\mbox{Pair occupation}}
\\[-5ex]
%%%%%
%%%%%
\botrule
\end{tabular}
\end{center}
\caption{\raggedright\small
The analogous states and operators in the correspondence between
the self interacting neutrinos and BCS pairing.
%Note that the quasi particle-hole operators of the pairing scheme are analogous
%to the neutrino flavor operators only in the strongly interacting limit.
}
\label{analogy_table_states}
\end{table}
%%%%%%%%%%%%%%%%%%%%%%%%%%%%%%%%%%%%%%%%%%%%%%%%%%%%%%%%%%%

%%%%%%%%%%%%%%%%%%%%%%%%%%%%%%%%%%%%%%%%%%%%%%%%%%%%%%%%%%%
\begin{table}[t]
\begin{center}
\begin{tabular}{r c c l} \toprule
\multicolumn{2}{c}{Self interacting neutrinos}
&
\multicolumn{2}{c}{Fermions with pairing}
\\
\hline
\\[-2ex]
%%%%%
\mbox{Oscillation frequency}
&
$\displaystyle{\omega_k}$
&
$\displaystyle{2\epsilon_k}$
&
\mbox{Pair energy}
\\[3ex]
%%%%%
\mbox{Split frequency}
&
$\displaystyle{\omega_c}$%(vacuum oscillation frequency)
&
$\displaystyle{2\epsilon_F}$
&
\mbox{Fermi energy}
\\[3ex]
%%%%%
\mbox{Lagrange multiplier}
&
$\lambda$
&
$2\mu$
&
\mbox{Chem. potential}
\\[3ex]
%%%%%
\mbox{Interaction strength}
&
$\displaystyle{G}$
&
$2g$
&
\mbox{Pairing strength}
\\[3ex]
%%%%%
\mbox{Neutrino mean field}
&
$\displaystyle{P^\pm}$
&
$\displaystyle{\Delta^\pm}$
&
\mbox{Pair mean field}
\\[3ex]
%%%%%
\mbox{Neutrino numbers}
&
\multirow{2}{*}{$\displaystyle{\frac{n_1\mkern-1mu(\mkern-1mu k \mkern-1mu) \mkern-3mu -
\mkern-3mu n_2 \mkern-1mu( \mkern-1mu k \mkern-1mu)}{2}}$}
&
\multirow{2}{*}{$\displaystyle{n_p \mkern-1mu ( \mkern-1mu k \mkern-1mu) \mkern-2mu - \mkern-2mu
\frac{1}{2}}$}
&
\mbox{Pair occupation}
\\[-1ex]
%%%%%
\mbox{in mass eigenstates}
&
&
&
\mbox{number}
\\[3ex]
%%%%%
%%%%%
\botrule
\end{tabular}
\end{center}
\caption{\raggedright\small
The list of analogous scalar quantities in the correspondence between
the self interacting neutrinos and BCS pairing.
The last two lines
follow from the analogy between pair quasispin and neutrino isospin, i.e.,
Eqs.\s(\ref{quasispin}) and (\ref{isospin}). Here $n_i(k)$ is the number of
neutrinos in the eigenstate with mass $m_i$ in the $k^{\mbox{\tiny th}}$
energy mode and $n_p(k)$ is the number of pairs in the $k^{\mbox{\tiny
th}}$ energy level.}
\label{analogy_table_scalar}
\end{table}
%%%%%%%%%%%%%%%%%%%%%%%%%%%%%%%%%%%%%%%%%%%%%%%%%%%%%%%%%%%

%%%%%%%%%%%%%%%%%%%%%%%%%%%%%%%%%%%%%%%%%%%%%%%%%%%%%%%%%%%%%%%%%%%%%%%%%%%
\section{BCS-BEC crossover and spectral splits}
The ground state of the pairing Hamiltonian evolves from
weakly bound Cooper pairs in the BCS limit of vanishing interactions to
the Bose-Einstein condensation of tightly bound diatomic molecules in the limit
of strong interactions. This evolution takes place without a phase transition as the
interaction strength is varied and hence the ground state can be described by the same variational BCS wave function
throughout the crossover between BCS and BEC limits~\cite{PhysRev.186.456,Leggett80}.
The theoretical prediction has been experimentally observed in ultra cold atomic
systems~\cite{PhysRevLett.92.040403, PhysRevLett.92.120401, PhysRevLett.92.120403, PhysRevLett.92.150402, PhysRevLett.93.050401}.
%The BCS-BEC crossover is the evolution of weakly bound Cooper pairs in the BCS
%limit of vanishing interactions to the Bose-Einstein condensation of tightly bound diatomic molecules in
%the limit of strong interactions. The ground state evolves without a phase
%transition and hence can be described by the variational BCS wave function
%throughout crossover between the BCS and BEC limits~\cite{PhysRev.186.456,Leggett80}.
%The theoretical prediction has been experimentally observed in ultra cold atomic
%systems~\cite{PhysRevLett.92.040403, PhysRevLett.92.120401, PhysRevLett.92.120403, PhysRevLett.92.150402, PhysRevLett.93.050401}.

%%%%%%%%%%%%%%%%%%%%%%%%%%%%%%%%%%%%%%%%%%%%%%%%%%%%%%%%%%%
\begin{figure*}
\resizebox{0.95\textwidth}{!}{\includegraphics{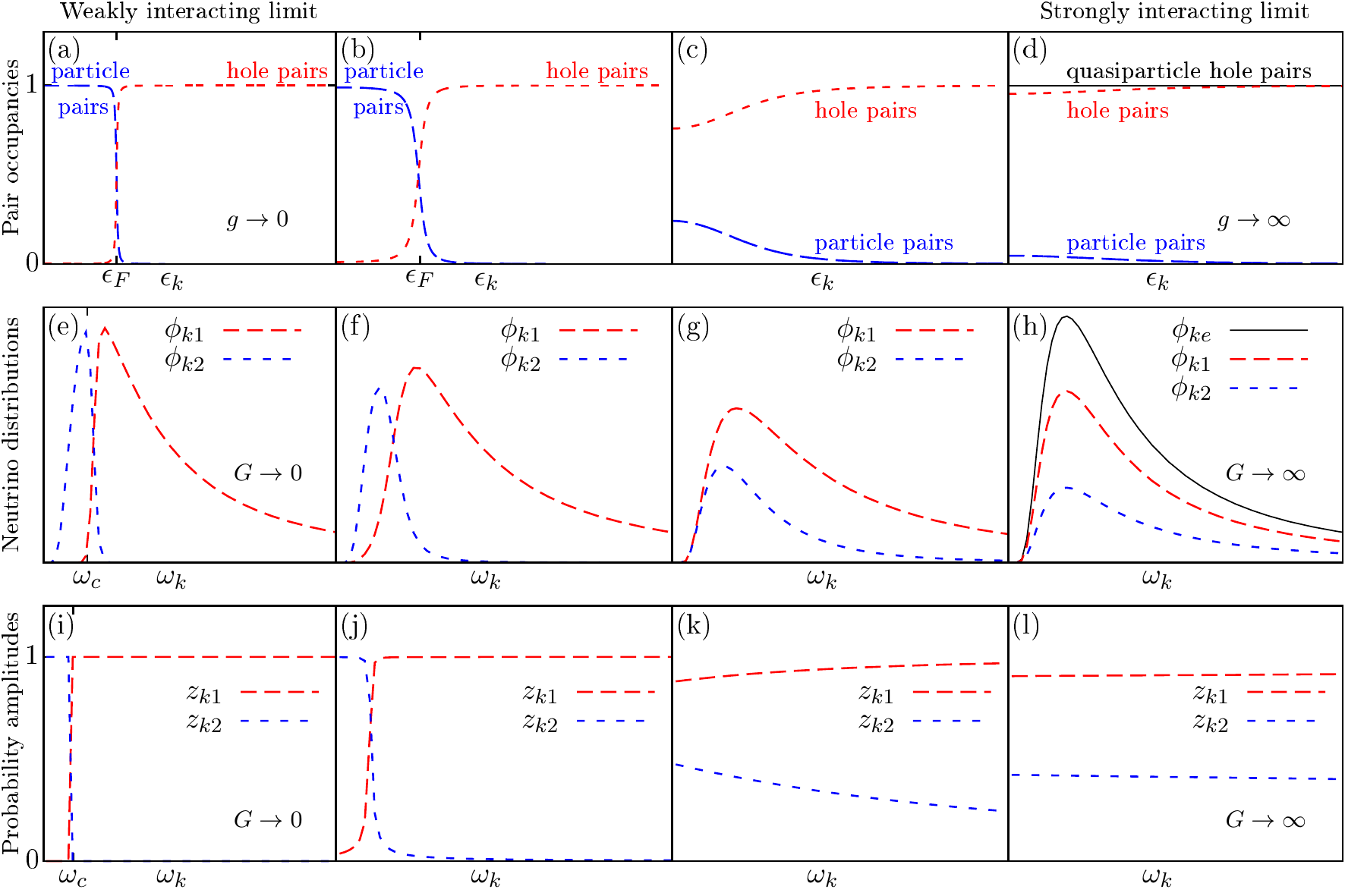}}
\vspace*{-0mm}
%\captionsetup{justification=RaggedRight}
\caption{\small
(Color online)
\emph{Upper panel:}
Evolution of pair distributions as the interaction strength increases from
left to right. (a) and (d) show the weakly (BCS) and strongly (BEC) interacting
limits of BCS-BEC crossover, respectively.
\emph{Middle panel:}
Evolution of neutrino distributions as the
neutrinos move away from the supernova core (strongly interacting regime) to the
outer layers (weakly interacting regime)
from right to left. (h) shows the assumed $\nu_e$ distribution
near the supernova core which is normalized to $1$, as well as the corresponding $\nu_1$ and $\nu_2$
distributions. As the neutrinos evolve
from strong to weak interaction regime, a spectral
split develops as shown on (e).
\emph{Lower panel:}
Evolution of the probability amplitudes $z_{1k}$ and $z_{2k}$, of finding a
neutrino in the first and second mass eigenstates, respectively.
}
\label{bcs-bec and split}
%\vspace*{-5mm}
\end{figure*}
%%%%%%%%%%%%%%%%%%%%%%%%%%%%%%%%%%%%%%%%%%%%%%%%%%%%%%%%%%%

The variational BCS ground state of the Hamiltonian in Eq.\s(\ref{cold atoms}) can be written as
\begin{eqnarray}
\label{BCS state}
|\mbox{\footnotesize BCS}\rangle &=& \prod_k \left(
\cos\theta_k + \sin\theta_k c_{k\uparrow}^\dagger c_{\bar{k}\downarrow}^\dagger
\right)|\emptyset\rangle \nonumber \\
&=& \prod_k \left(\cos\theta_k |\textrm{---}\rangle_k + \sin\theta_k |\f\rangle_k\right)
\end{eqnarray}
%\begin{equation}
%\label{BCS state}
%|\mbox{\footnotesize BCS}\rangle=\prod_k \left(
%\cos\theta_k + \sin\theta_k c_{k\uparrow}^\dagger c_{\bar{k}\downarrow}^\dagger
%\right)|\emptyset\rangle\msp,
%\end{equation}
where the angle $\theta_k$ is to be found from
\begin{equation}
\label{theta_k}
\cos{2\theta_k}=\frac{(\epsilon_k-\mu) }{E_k}
\end{equation}
with
\begin{equation}
\label{E_k}
E_k=\sqrt{(\epsilon_k - \mu)^2 + g^2\Delta^+\Delta^-}.
\end{equation}
%\begin{equation}
%\label{BCS angle}
%\cos{2\theta_k}=\frac{\epsilon_k - \mu }{\sqrt{(\epsilon_k - \mu)^2 + g^2\Delta^+\Delta^-}}\msp.
%\end{equation}
The requirement that the state in Eq.\s(\ref{BCS state}) should satisfy Eq.\s(\ref{mean field pairs})
gives rise to the self-consistency equations
\begin{equation}
\label{self-consistency}
\begin{split}
	\frac{1}{g}&=\sum_k \frac{f_k-\frac{1}{2}}{E_k},\\
	\sum_k \left(n_p(k)-\frac{1}{2}\right)&=\sum_k\frac{\epsilon_k-\mu}{E_k}\left(f_k-\frac{1}{2} \right)
\end{split}
\end{equation}
which determine the mean field $\Delta^\pm$ and the chemical potential $\mu$.
%Here $n_p(k)$ is the number of Cooper pairs in the $k^{\mbox{\tiny th}}$ energy level.
The distribution of the quasiparticles is given by the Fermi function
\begin{equation}
f_k=\frac{1}{\exp(E_k/k_B T)+1}
\end{equation}
where $T$ is the temperature and $k_B$ is the Boltzmann constant.
It is also useful to define the quasiparticle operators
\begin{equation}
\label{quasiparticles}
\tilde{c}_{k\uparrow}^{\phantom{\dagger}}
\mkern-6mu=\mkern-4mu\cos{\theta_k} c_{k\uparrow}^{\phantom{\dagger}}
\mkern-6mu-\mkern-1mu \sin{\theta_k} c_{\bar{k}\downarrow}^\dagger\msp,
\mkern14mu
\tilde{c}_{\bar{k}\downarrow}^{\phantom{\dagger}}
\mkern-6mu=\mkern-4mu\sin{\theta_k} c_{k\uparrow}^\dagger
\mkern-6mu+\mkern-1mu \cos{\theta_k} c_{\bar{k}\downarrow}^{\phantom{\dagger}}\msp,
\end{equation}
which annihilate the BCS ground state so that $|\mbox{\footnotesize BCS}\rangle$
can be viewed as a quasiparticle vacuum.
Accordingly, pair occupation numbers in the $|\mbox{\footnotesize BCS}\rangle$ state are given by
\begin{equation}
\label{pair occupations}
n_p(k)=\langle c_{k\uparrow}^\dagger c_{k\uparrow}^{\phantom{\dagger}} \rangle =
\langle c_{\bar{k}\downarrow}^\dagger c_{\bar{k}\downarrow}^{\phantom{\dagger}} \rangle
= \sin^2\theta_k\msp.
\end{equation}
These occupation numbers can be calculated for any value of the interaction
constant $g$ by first solving the self consistency equations given in Eq.
(\ref{self-consistency}) for the mean field $\Delta^\pm$ and chemical potential
$\mu$, and then calculating $\theta_k$ from Eqs.\s(\ref{theta_k}) and
(\ref{E_k}).

In the weak interaction limit, the solution describes the non-interacting Fermi sea with
\begin{equation}
\label{g 0}
\lim_{g\to 0} \theta_k=\frac{\pi}{2}\Theta(\epsilon_F-\epsilon_k)
\end{equation}
%\begin{equation}
%\label{g 0}
%\lim_{g\to 0} \theta_k=
%\begin{cases}
%\tfrac{\pi}{2} & \mbox{ for } \epsilon_k<\epsilon_F\msp, \\
%0 & \mbox{ for } \epsilon_k>\epsilon_F\msp.
%\end{cases}
%\end{equation}
where $\Theta(x)$ denotes the Heaviside step function and
\begin{equation}
\lim_{g\to 0} \mu=\epsilon_F
\end{equation}
is the Fermi energy. Fig.~\ref{bcs-bec and split}a 
shows the corresponding particle (blue dashed line) and hole (red
dotted line) occupation numbers in this limit which follow from substituting
Eq.\s(\ref{g 0}) in Eq.\s(\ref{pair occupations}). The particle pairs fill the
levels up to the Fermi energy and the system displays the characteristic
distribution of a degenerate ideal Fermi gas at zero temperature.
%The quasiparticle holes which make
%up the $|\mbox{\footnotesize BCS}\rangle$ state are more particle-like for
%$\epsilon_k<\epsilon_F$ and more hole-like for $\epsilon_k>\epsilon_F$ in
%this limit.

%As one moves from left to right in the upper panel of Fig.~\ref{bcs-bec and
%split}, the interaction strength increases and the distributions are gradually smoothed
The interaction strength increases from left to right in the upper panel of Fig.~\ref{bcs-bec and
split} and the distributions are gradually smoothed
out as more and more levels start to take part in pairing.
In the limit of strong
interactions the angle $\theta_k$ tends to the same value for all pairs, i.e.,
\begin{equation}
\lim_{g\to\infty}{\theta_k}=\theta\msp.
\end{equation}
In this limit, $|\mbox{\footnotesize BCS}\rangle$ represents a BEC in the form of a
coherent state of atomic pairs occupying the same single-pair quantum state.
Fig.~\ref{bcs-bec and split}d displays the occupation
numbers of particle and hole pairs which are almost uniform in this limit
indicating that all levels are taking part in pairing.  Here, quasihole
distribution is also shown with a solid black line.  Note that the quasihole
distribution is equal to unity and remains the same throughout the crossover for
any value of $g$ because $|\mbox{\footnotesize BCS}\rangle$ is the quasiparticle
vacuum, but it is indicated only in this plot to emphasize its
resemblence to the $\nu_e$ distribution in the strong neutrino self interaction
regime (see below).

\begin{figure}[!t]
%\resizebox{\columnwidth}{!}{\input{plot_chemical_potential.tex}}
\resizebox{\columnwidth}{!}{\includegraphics{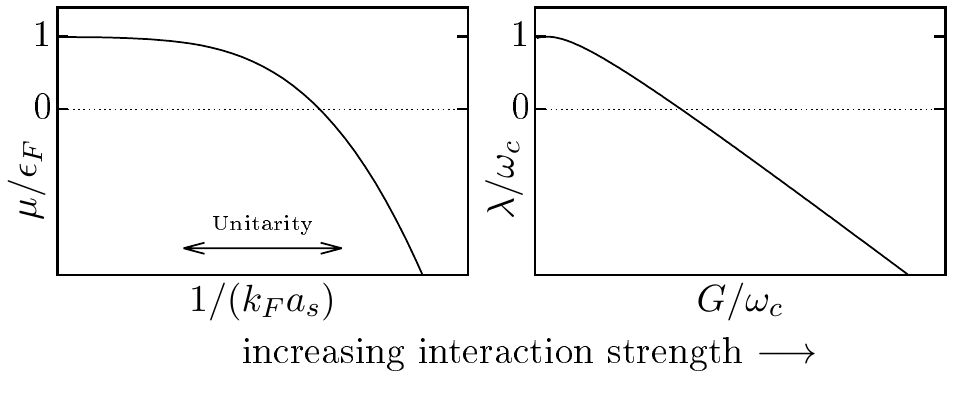}}
%\vspace*{5mm}
%\captionsetup{justification=RaggedRight}
\caption{\small
\emph{Left panel:} Chemical potential $\mu$ calculated from Eqs.\s(\ref{mean field pairs})
%calculated self-consistently from Eqs.\s(\ref{mean field pairs}) and plotted against the inverse of the scattering length $a_s$.
vs. the scattering length $a_s$ characterizing the interactions.
($\epsilon_F$/$k_F$ denotes the Fermi energy/momentum.)% $1/(k_Fa_s)$ characterizes the interaction strength $g$.
%\emph{Right panel:} Lagrange multiplier $\lambda$ calculated self-consistently from Eqs.\s(\ref{mean field neutrinos})
\emph{Right panel:} Lagrange multiplier $\lambda$ calculated from Eqs.\s(\ref{mean field neutrinos})
vs. the interaction parameter $G$. ($\omega_c=\lim_{G\to 0} \lambda$)
For both panels, the analogous parameters $g$ and $G$ increase from left to
right. $\mu=0$ and $\lambda=0$ are the boundary points between BCS and BEC
regimes.
\label{chemical potential}}
%\vspace*{-5mm}
\end{figure}

The typical evolution of the chemical potential in the BCS-BEC crossover
from positive values to negative values is shown in the left panel of
Fig.~\ref{chemical potential} as a function of the inverse scattering length,
the parameter characterizing the strength of the pairing interaction.  The
vanishing of the chemical potential is accompanied with the shift of excitation
energy minimum to zero momentum and is identified as the separation point
between the BCS and BEC sides of the crossover.

For self interacting neutrinos, the state which is analogous to
$|\mbox{\footnotesize BCS}\rangle$  can be written down using Table~\ref{analogy_table_states}:
\begin{eqnarray}
\label{BCS state neutrino}
|\mbox{\footnotesize ``BCS''}\rangle &=& \prod_k \left(
\cos\theta_k + \sin\theta_k a_{k2}^\dagger a_{k1}^{\phantom{\dagger}}
\right) |\nu_1 \; \nu_1 \; \nu_1 \; \dots \rangle \nonumber \\
&=& \prod_k \left(\cos\theta_k |\nu_1\rangle_k + \sin\theta_k |\nu_2\rangle_k\right)
\end{eqnarray}
Here, the angle $\theta_k$ and the associated self consistency equations are the
same as those given in Eqs.\s(\ref{theta_k}-\ref{self-consistency}) with the
replacements shown in Table (\ref{analogy_table_scalar}) and $f_k-1/2 \to
-\phi_e/2$ where $\phi_e$ is the Fermi function describing electron neutrino
energy distribution. The last replacement follows from the analogy between pair
quasispin and neutrino isospin (see Eqs.\s(\ref{quasispin}) and (\ref{isospin})).
Due to the overall sign difference between the two Hamiltonians given in
Eq.\s(\ref{hamiltonians}), $|\mbox{\footnotesize ``BCS''}\rangle$ is not the
ground state of the neutrino Hamiltonian, but its highest energy eigenstate.
However, since the energy spectra of the two Hamiltonians are the same apart
from an overall sign, $|\mbox{\footnotesize ``BCS''}\rangle$ should also evolve
smoothly between strong and weak interaction regimes without a phase transition
(i.e., with no level crossings).

Using Eq.\s(\ref{mapping}), one can also define the analogs of the quasiparticle
operators introduced in Eq.\s(\ref{quasiparticles})
\begin{equation}
\label{quasiparticles for neutrinos}
\tilde{a}_{k2}^{\phantom{\dagger}}\mkern-6mu=\mkern-4mu
\cos{\theta_k} a_{k2}^{\phantom{\dagger}} \mkern-5mu - \mkern-1mu \sin{\theta_k} a_{k1}\msp,
\mkern7mu
\tilde{a}_{k1}^\dagger \mkern-6mu=\mkern-4mu
\sin{\theta_k} a_{k2}^\dagger \mkern-4mu + \mkern-1mu \cos{\theta_k} a_{k1}^\dagger\msp,
\end{equation}
which similarly annihilate the $|\mbox{\footnotesize ``BCS''}\rangle$ state.
Note that, unlike the operators in Eq.\s(\ref{quasiparticles}), these operators
do not mix particle and hole states which is consistent with the number
conserving nature of the neutrino self interactions. Denoting the states
associated with the operators $\tilde{a}_{1}^\dagger$ and
$\tilde{a}_{2}^\dagger$ by $|\tilde{\nu}_1\rangle$ and
$|\tilde{\nu}_2\rangle$, respectively, this tells us that the
$|\mbox{\footnotesize ``BCS''}\rangle$ state is a $|\tilde{\nu}_1\rangle$
condensate.

For neutrinos, the limit of strong self interactions is realized near
the core of the supernova. In general, the flavor composition of neutrinos
released from the core depends on the explosion phase. Here we consider a
special case in which all neutrinos are released as $\nu_e$, which is
more relevant for the initial stages of the explosion. For such
a configuration, the solution of the
consistency equations given in Eq.\s(\ref{mean field neutrinos}) yield
\begin{equation}
\lim_{G\to\infty}\theta_k=\vartheta
\end{equation}
where $\vartheta$ is the neutrino mixing
angle introduced in Eq.\s(\ref{neutrino mixing}). This is similar to the BEC
regime of the fermion pairs.
Substituting this angle in Eq.\s(\ref{BCS state neutrino}) and using Eq.\s(\ref{neutrino mixing}) gives
\begin{equation}
|\mbox{\footnotesize ``BCS''}\rangle= |\nu_e \; \nu_e \; \nu_e \; \dots \rangle
\end{equation}
which confirms the self-consistency of the state. In this
limit, the quasiparticle operators which annihilate this state become the particle operators in
flavor basis, i.e.,
\begin{equation}
%\label{}
\tilde{a}_{k2}^{\phantom{\dagger}} = a_{k\mu}^{\phantom{\dagger}}
\quad\mbox{and}\quad
\tilde{a}_{k1}^\dagger = a_{ke}^\dagger\msp.
\end{equation}
In Fig.~\ref{bcs-bec and split}h, we plot
neutrino occupation numbers associated with the $|\mbox{\footnotesize ``BCS''}\rangle$
state in this limit. Note that, although the pairing Hamiltonian in
Eq.\s(\ref{cold atoms}) describes the atomic pairs at ultra low temperatures,
the Hamiltonian in Eq.\s(\ref{neutrinos}) represents self interactions of
neutrinos for any (thermal or non-thermal) energy distribution. The main features
of the analogy is independent of the neutrino energy distribution. For
illustration, we use a thermal $\nu_e$ distribution with a temperature of $5$
MeV which is shown with the solid-black line. The corresponding $\nu_1$ and $\nu_2$
occupation numbers which follow from Eq.\s(\ref{neutrino mixing}) are shown with
red-dotted and blue-dashed lines, respectively.

%%%%%%%%%%%%%%%%%%%%%%%%%%%%%%%%%%%%%%%%%%%%%%%%%%%%%%%%%%%
%%%%%%%%%%%%%%%%%%%%%%%%%%%%%%%%%%%%%%%%%%%%%%%%%%%%%%%%%%%

As the neutrinos move away from the core of the supernova, $\nu\nu$ interactions
are gradually turned off as described by Eq. (\ref{G}). Therefore, one expects
the $|\mbox{\footnotesize ``BCS''}\rangle$ state to evolve in a way which is
similar to the BCS-BEC crossover of $|\mbox{\footnotesize BCS}\rangle$ state,
but reversed in the direction of decreasing interaction strength.
The middle panel of Fig.~\ref{bcs-bec and split} shows
the evolution of the neutrino distributions as the neutrino self interaction
constant $G$ decreases from \emph{right to left}. In the dilute, weakly
interacting regime where
\begin{equation}
\lim_{G\to 0}\lambda=\omega_c\msp,
\end{equation}
the state $|\mbox{\footnotesize ``BCS''}\rangle$ in Eq. (\ref{BCS state
neutrino}) evolves into
\begin{equation}
|\mbox{\footnotesize ``BCS''}\rangle = \prod_{\omega_k<\omega_c}|\nu_2\rangle_k
\prod_{\omega_k>\omega_c}|\nu_1\rangle_k
\end{equation}
under the adiabatic evolution conditions. This distribution is plotted in 
Fig.~\ref{bcs-bec and split}e and corresponds to the BCS
limit of the fermion pairing. This
is a particular example of a \emph{spectral split}
phenomenon, so called because the original $\nu_e$ energy distribution is
eventually split between the two mass eigenstates. This phenomenon was observed
in numerical simulations of supernova neutrinos by various groups (see Refs.
\cite{Duan:2010bg,Mirizzi:2015eza} for review).

The similarity between pair distribution of cold atoms which we treat at zero
temperature, and the neutrino distribution which we treat at finite temperature
becomes pronounced if we focus on the \emph{Bogoliubov coefficients}
\begin{equation}
\label{z_k}
z_{k1}=\frac{\langle a_{k1}^\dagger \tilde{a}_{k1}\rangle}{\langle
\tilde{a}_{k1}^\dagger \tilde{a}_{k1}\rangle}=\cos{\theta_k}\msp,
\mkern5mu
z_{k2}=\frac{\langle a_{k2}^\dagger \tilde{a}_{k1}\rangle}{\langle
\tilde{a}_{k1}^\dagger \tilde{a}_{k1}\rangle}=\sin{\theta_k}\msp.
\end{equation}
These are the probability amplitutes for the neutrino born in the state
$|\nu_e\rangle$ near the core (where it almost overlaps with
$|\tilde{\nu}_1\rangle$) to be found in $|\nu_1\rangle$ or $|\nu_2\rangle$ mass
eigenstate, respectively.
The evolutions of these coefficients are shown in the lower panel of Fig.~\ref{bcs-bec and split}
as a function of the interaction constant.
In strongly interacting limit near the center of the
supernova Bogoliubov coefficients are uniform with $z_{k1}\to\cos\vartheta$ and
$z_{k2}\to\sin\vartheta$ but as the neutrinos move away from the center,
$|\tilde{\nu}_{1}\rangle$ becomes more and more like $|\nu_2\rangle$ (or
$|\nu_1\rangle$) for low (high) $\omega$ values.  This is reminiscent of the fact
that, during the BCS-BEC crossover, quasihole degrees of freedom at the BEC
limit coincide with real particles (holes) for low (high) energies at the BCS
limit.

In Fig.~\ref{gap}, we plot the eigenvalues of the fermion pairing (left panel)
and self interacting neutrino (right panel) Hamiltonians for three
representative values of the respective interaction strengths. For the fermion
pairs, these eigenvalues are $\pm E_k$ with $E_k$ given by Eq.~(\ref{E_k}). For
neutrinos, they are given by the same formula with the replacements from Table
\ref{analogy_table_scalar}. The solid black lines represent the weakly
interacting limits where the chemical potential $\mu$ and the Lagrange
multiplier $\lambda$ almost coincide with their limiting values which are the
Fermi energy $\epsilon_F$, and the split frequency $\omega_c$, respectively. The
dashed blue lines correspond to the point at which $\mu$ and $\lambda$ become
zero, and the dotted red lines represent the regime in which they are negative.
In the case of fermion pairs, the difference between these eigenvalues ($2E_k$)
is the energy gap for the creation of a quasiparticle pair excitation in the
system. This energy gap is minimized  at $\epsilon_k=\mu$ on the BCS side, i.e. while $\mu>0$, 
and at $\epsilon_k=0$ on the BEC side, i.e. while $\mu<0$.
As a result,
as the interaction constant increases,
the location of the minimum moves from the Fermi surface to the zero momentum. 
Therefore, the excitations from the BCS groundstate are always gapped.
In the case of neutrinos, the difference $2E_k$ is
the energy gap between the states $|\tilde{\nu}_1\rangle$ and
$|\tilde{\nu}_2\rangle$, which leads to the avoided level crossing
in this model. It is minimized for neutrinos with oscillation
frequency $\omega_k=\lambda$ while $\lambda>0$ and $\omega_k=0$ while $\lambda<0$. 
{Near the core of the supernova where self interactions dominate,
this minimum is located at $\omega_k=0$. As the
interaction strength decreases, the minimum gap moves until it
eventually reaches to $\omega_c$.

For the cold atoms, the shifting of the location of the  minimum gap to zero
momentum with increasing $g$ occurs when $\mu=0$ which indicates that the system
has moved from BCS to BEC side.  For the neutrinos, although $G$ decreases as
they move away from the supernova core, the energy distributions initially do
not change significantly. The change begins soon after  the location of the
minimum gap moves to $\omega=0$ which occurs when $\lambda=0$. Therefore, the
fact that the minimum gap coincides with $\omega=0$ can be seen as the beginning
of the split phenomenon.

%%%%%%%%%%%%%%%%%%%%%%%%%%%%%%%%%%%%%%%%%%%%%%%%%%%%%%%%%%%
\begin{figure}
\resizebox{\columnwidth}{!}{\includegraphics{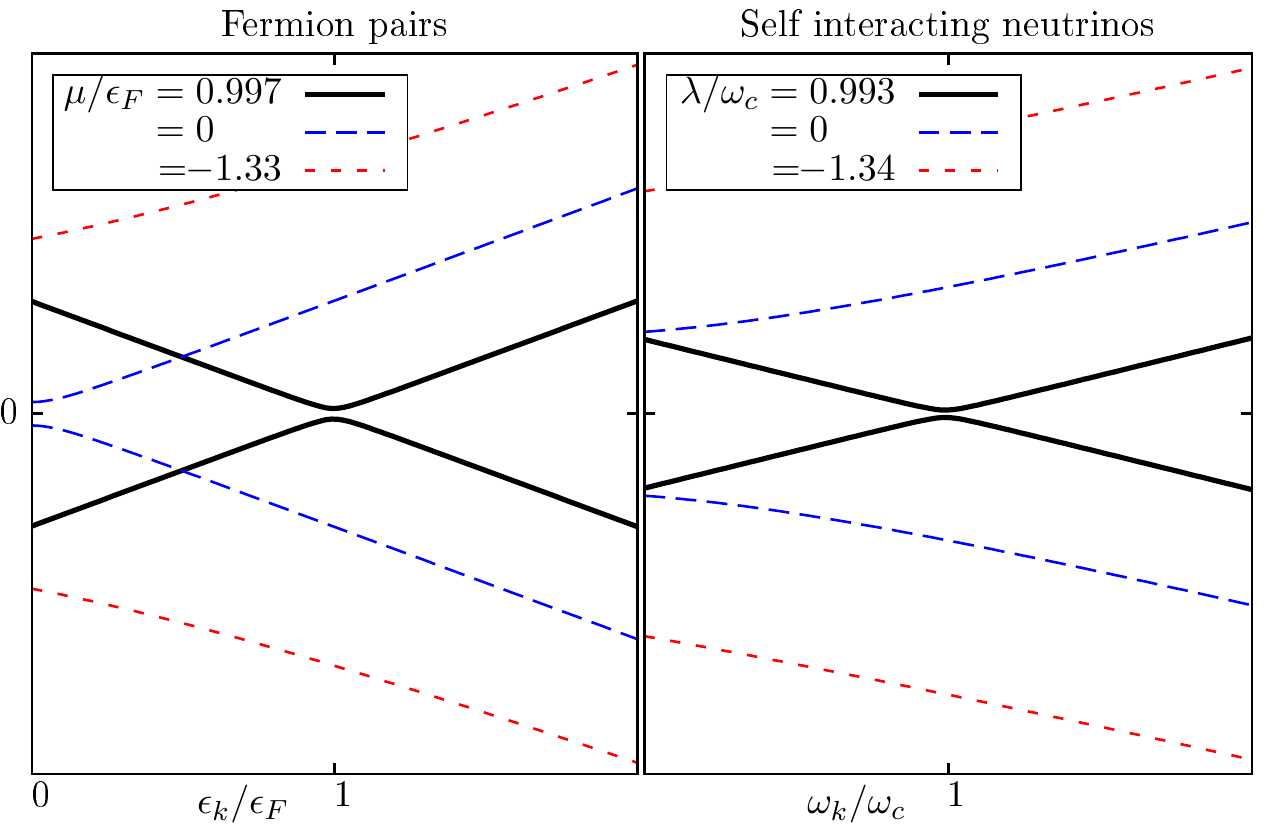}}
%\vspace*{-5mm}
%\captionsetup{justification=RaggedRight}
\caption{\small
\emph{Left panel:} Eigenvalues of the Hamiltonian matrix for fermion pairs given
in Eq. (\ref{cold atoms}) at three different values of the interaction constant
$g$ corresponding to three different values of the chemical potential.
$\mu/\epsilon_F=1$ (solid black line) is on the BCS limit, 
$\mu/\epsilon_F<0$ is on BEC side (red dotted line), and
$\mu=0$ (blue dashed line) is the boundary between BCS and BEC regimes.
\emph{Right panel:} Eigenvalues of the Hamiltonian matrix for self interacting
neutrinos given in Eq. (\ref{neutrinos})
for three different values of the interaction constant
$G$ corresponding to positive, zero, and negative values of the Lagrange
multiplier $\lambda$. $\lambda=0$ is the boundary between BCS and BEC
regimes.
}
\label{gap}
\end{figure}
%%%%%%%%%%%%%%%%%%%%%%%%%%%%%%%%%%%%%%%%%%%%%%%%%%%%%%%%%%%

\section{Summary \& Conclusions}
We showed an analogy between the spectral split of a neutrino ensemble which
initially consists of electron type neutrinos, and the BCS-BEC
crossover phenomenon. This analogy is illustrated using the BCS model of ultra cold atomic gases.  Due to
their mathematically equivalent Hamiltonians, these systems undergo identical
transformations while they evolve from strong to weak interaction regimes.  In
particular, the Fermi energy of the BCS model and the critical split frequency
of neutrinos play analogous roles.  Note that although the pair quasispin and
the neutrino isospin obey the same algebra, the latter is number conserving
whereas the former is not. As a result, in Eqs.\s(\ref{vacuum}-\ref{others}),
BCS pair states live in a Fock space whereas the neutrino states live in a
regular Hilbert space. The energy gap plotted in Fig.~\ref{gap} is a
quasipair excitation gap in the case of cold atoms from the ground state,
whereas for neutrinos, it is the energy difference between the two mixing
eigenstates.

In the BEC limit, all atomic pairs occupy the same state so that the
configuration is maximally symmetric.
The analog neutrino state should also be maximally symmetric, i.e., all
neutrinos should initially have the same flavor.  We choose that to be $\nu_e$
to associate the initial state with the neutralization burst of supernova, but
any other flavor could have been chosen. Antineutrinos of the opposite flavor
($\bar{\nu}_\mu$ in our case) could also be added to the picture without
breaking the analogy because neutrinos and antineutrinos transform under the
conjugate representations, e.g., $\bar{\nu}_\mu$ transforms in the same way as
$\nu_e$. In other words, the BCS-BEC crossover analogy considered here would hold for
any maximally symmetric initial state which consists of neutrinos of one type, and
antineutrinos of the opposite type. For such a state, the split occurs in the
neutrino sector and the behavior of antineutrinos is dictated by the
\emph{lepton number conservation}, i.e., the conservation of $P^0$ defined in
Eq. (\ref{mean field neutrinos}) (\cite{Duan:2006an,Fogli:2007bk}).  Other
neutrino spectral split scenarios involving less symmetric initial
configurations,  do not correspond to a simple BCS-BEC crossover. Such initial
neutrino states may display more complicated behavior including multiple
spectral splits~\cite{Dasgupta:2009mg}.  It is an open question weather or not
these splits correspond to some other phenomena in the fermion pairing scheme.

In a realistic supernova setting small quantities of non-electron flavor
neutrinos would be present in the initial deleptonization phase.  Moreover,
recent simulations suggest that the neutrino spectral splits are unstable
against the inclusion of the multi angle and three flavor
effects~\cite{Friedland:2010sc,Raffelt:2013rqa}, both of which are important in
a real supernova but are omitted in this study. Still, the present analogy can be
helpful in understanding some aspects of collective flavor oscillations in
supernova. For example, an experimental cold atom system can be used to
\emph{simulate} the possible contribution of entangled many-body states to the
collective behaviour of neutrinos~\cite{Friedland:2003dv, Friedland:2003eh,
Volpe:2013jgr,Pehlivan:2011hp,Pehlivan:2014zua}. Such a contribution will
present itself as a deviation of the experimental cold atom system from the
results obtained by the mean field approximation which is currently employed by
most numerical studies of supernova neutrinos, including the one presented here.
Possible departures from adiabaticity~\cite{Dasgupta:2010cd}, the factors
affecting the split frequency~\cite{Fogli:2009rd,Sarikas:2012ad,Raffelt:2011yb},
and even the multi angle instability of split behavior mentioned above can be subjects of such
an experimental study. Moreover, in the full three flavor mixing case neutrino isospin generalizes to an $SU(3)$
operator~\cite{Pehlivan:2014zua}. There are pairing scenarios for atomic
systems~\cite{PhysRevB.70.094521, Bedaque20091763} and in QCD~\cite{RevModPhys.80.1455}
suggesting similar analogies in this case. A possible extension of our analogy
in this direction may help us to gain insight about the three flavor instability.

Finally, other quantum many-body systems, in which the relative strength of kinetic
and interaction energies is density dependent, might also have been considered
in lieu of ultra cold atomic Fermi gases in our discussion.  For instance, in
the context of excitonic condensates, the BCS-BEC crossover is driven by density~\cite{Andrenacci99}.
In such electronic systems with long range Coulomb interactions, somewhat
counter-intuitively, the low density limit results in a strongly interacting
system. For neutrinos, the self interaction term is density dependent because
many scattering amplitudes must coherently superpose to generate the effect.
Thus, unlike the case of excitonic condensates, the relative strength of
neutrino self interactions decreases with density.

\vspace*{3mm} \noindent Y.P. thanks to CETUP* 2015 organizers for allowing a
stimulating environment. This work was supported
%Yamac
by T{\"{U}}B{\.{I}}TAK under project number 115F214.

\bibliography{apj-jour,biblio_yamac}

\end{document}